\documentclass{article}
   \usepackage{graphicx}
   \usepackage{apjgalley}
 \usepackage{multicol}
\usepackage{psfig}
  \font\sevenrm=cmr7

\def\prl{Phys. Rev. Letters} 
\def\app{Astroparticle Phys.}  
\def\mnras{{Mon. Not. R. Ast. Soc.}} 
\def\ssr{Space Sci. Rev.}  
\def\teq#1{$\, #1\,$} 
\def\etal{\it et al. \rm\hskip 0.1em }

\def\thetascatt{\theta_{\hbox{\sevenrm scatt}}}
\def\thetaBone{\Theta_{\hbox{\sevenrm B1}}}

\def\today{\ifcase\month\or
  January\or  February\or  March\or  April\or May\or  June\or  July\or
August\or September\or October\or November\or December\fi
\space\number\day, \number\year}
\begin{document}
%
%
\newcommand{\vol}[2]{$\,$\rm #1\rm , #2.}  
\newcommand{\figureout}[5]{\centerline{}
   \centerline{\hskip #3in \psfig{figure=#1,width=#2in}} \vspace{#4in}
   \figcaption{#5} }
\newcommand{\twofigureout}[3]{\centerline{}
   \centerline{\psfig{figure=#1,width=3.4in} \hskip 0.5truein
\psfig{figure=#2,width=3.4in}} \figcaption{#3} } 
\newcommand{\figureoutpdf}[5]{\centerline{} \vskip -5pt 
   \centerline{\hspace{#3in} \includegraphics[width=#2truein]{#1}}
\vspace{#4truein} \figcaption{#5} \centerline{} }
\newcommand{\twofigureoutpdf}[3]{\centerline{}
   \centerline{\includegraphics[width=3.4truein]{#1}
        \hspace{0.5truein} \includegraphics[width=3.4truein]{#2}}
\vspace{-0.2truein} \figcaption{#3} } 
%
\newcommand{\tableoutpdf}[5]{\centerline{}
  \vspace{#3truein} \centerline{\hspace{#4in}
\includegraphics[width=#2truein]{#1}}
 \vspace{#5truein}
\centerline{} }
\newcommand{\figureoutsubmit}[5]{
   \begin{figure*} \plotone{#1} \vspace{#4truein} \caption{#5}
\end{figure*} } 
\newcommand{\twofigureoutsubmit}[3]{
   \begin{figure*} \plottwo{#1}{#2} \caption{#3} \end{figure*} }
   
\submitted{Submitted to ApJ Letters June 26, 2007}
\title{BLAZAR GAMMA-RAYS, SHOCK ACCELERATION, AND THE \\
EXTRAGALACTIC BACKGROUND LIGHT}

\author{Floyd W. Stecker}
\affil{Astrophysics Science Division, NASA Goddard Space Flight Center \\
Greenbelt, MD 20771, U.S.A.\\
\it stecker@milkyway.gsfc.nasa.gov\rm} 

\smallskip

\author{Matthew G. Baring and Errol J. Summerlin}
   \affil{Department  of   Physics  and  Astronomy   MS-108,  \\  Rice
University,  P.O.    Box  1892,   Houston,  TX  77251,   U.S.A.\\  \it
baring@rice.edu, xerex@rice.edu\rm}
%
%

\begin{abstract} 
The  observed spectra of  blazars, their  intrinsic emission,  and the
underlying populations of  radiating particles are intimately related.
The  use of  these sources  as  probes of  the extragalactic  infrared
background,  a  prospect  propelled  by recent  advances  in  TeV-band
telescopes, soon  to be augmented  by observations by  NASA's upcoming
{\it Gamma-Ray Large  Area Space Telescope (GLAST)}, has  been a topic
of great recent interest.  Here,  it is demonstrated that if particles
in   blazar  jets  are   accelerated  at
relativistic   shocks,  then
$\gamma$-ray spectra with indices less than 1.5 can be produced. This,
in turn, loosens  the upper limits on the  near infrared extragalactic
background  radiation  previously  proposed.   We also  show  evidence
hinting that  TeV blazars with  flatter spectra have  higher intrinsic
TeV  $\gamma$-ray luminosities  and we  indicate that  there may  be a
correlation of flatness and luminosity with redshift.
\end{abstract}
\section{INTRODUCTION}
 \label{sec:intro}
The  comparison of  theoretical  models for  the $\gamma$-ray  spectra
of
blazars  with  observations  is  the  standard  approach  used  to
understand
the  physical  processes  leading  to  their  high  energy
emission.   In  the
case  of  TeV components,  the  radiation  models
generally   involve  the
synchrotron   self-Compton   (SSC)  process.
Determining  the  intrinsic
emission  spectra  of  such  TeV  sources
requires  that  one account  for  the
energy  and redshift  dependent
absorption of $\gamma$-rays from these
sources through
\teq{\gamma\gamma\to e^+e^-} interactions with intergalactic photon
backgrounds  produced by  stellar and  dust  emission. The
cumulative
background radiation seen at redshift  $z = 0$ is commonly referred to
as  the  extragalactic background  light  (EBL). Various  calculations
of
extragalactic  $\gamma$-ray  absorption  have  been given  in  the
recent
literature  and  they  are  discussed along  with  the  latest
calculations  in
the  paper  of  Stecker,  Malkan  \&  Scully  (2006,
hereafter  SMS06).

The results  given  by SMS06  are  based on  two
galaxy evolution models,
{\it  viz.} a baseline model (B)  and a fast
evolution  model  (FE).   The
spectral  energy distributions  of  the
extragalactic  background   light  for
these  models   are  shown  in
Figure~\ref{fig:EBL}. The FE model is favored by recent
{\it Spitzer}
observations  (Le Floc'h \etal  2005, Perez-Gonzalez  \etal
2005). It
provides a better description of the deep {\it Spitzer} number
counts
at   70   and   160$\mu$m   than   the   B   model.    However,   {\it
GALEX}
observations indicate  that the evolution of  UV radiation for
$0 <  z < 1$  may be somewhat  slower and more consistent  with
the B
model within errors (Schiminovich  \etal 2005). And the 24$\mu$m
{\it
Spitzer}  source  counts  are  closer  to  the B  model  than  the  FE
model.

Probing the EBL using blazar observations is contingent upon
an accurate
understanding of their emission spectra.  Aharonian \etal
(2006) have
argued that  intrinsic blazar spectra  must have spectral
indices
$\Gamma_{s} \ge 1.5$. They use this assumption, together with
{\it HESS} observations of the source
1ES 1101-232, to place an upper
limit  on  the  EBL  of  14
nWm$^{-2}$sr$^{-1}$ at  a  near  infrared
wavelength of  1.5 $\mu$m,
corresponding  to a frequency  of $2\times
10^{14}$ Hz. As can  be seen
from both Figure~\ref{fig:EBL} and Table
1, this value is consistent with  model B,
but not with the model FE,
which is favored by  the {\it Spitzer}
observations.  It is therefore
important for exploring both galaxy
evolution and blazar physics that
we reexamine the assumption
$\Gamma_{s} \ge 1.5$ made by Aharonian
\etal (2006). This assumption has
been questioned by Katarz'{n}ski
\etal (2006) in a different context,
but we will examine it here in
the  light  of  the  physics of  shock
acceleration,  which  provides
insights    into    the    distributions    of
underlying    particle
populations.

\figureout{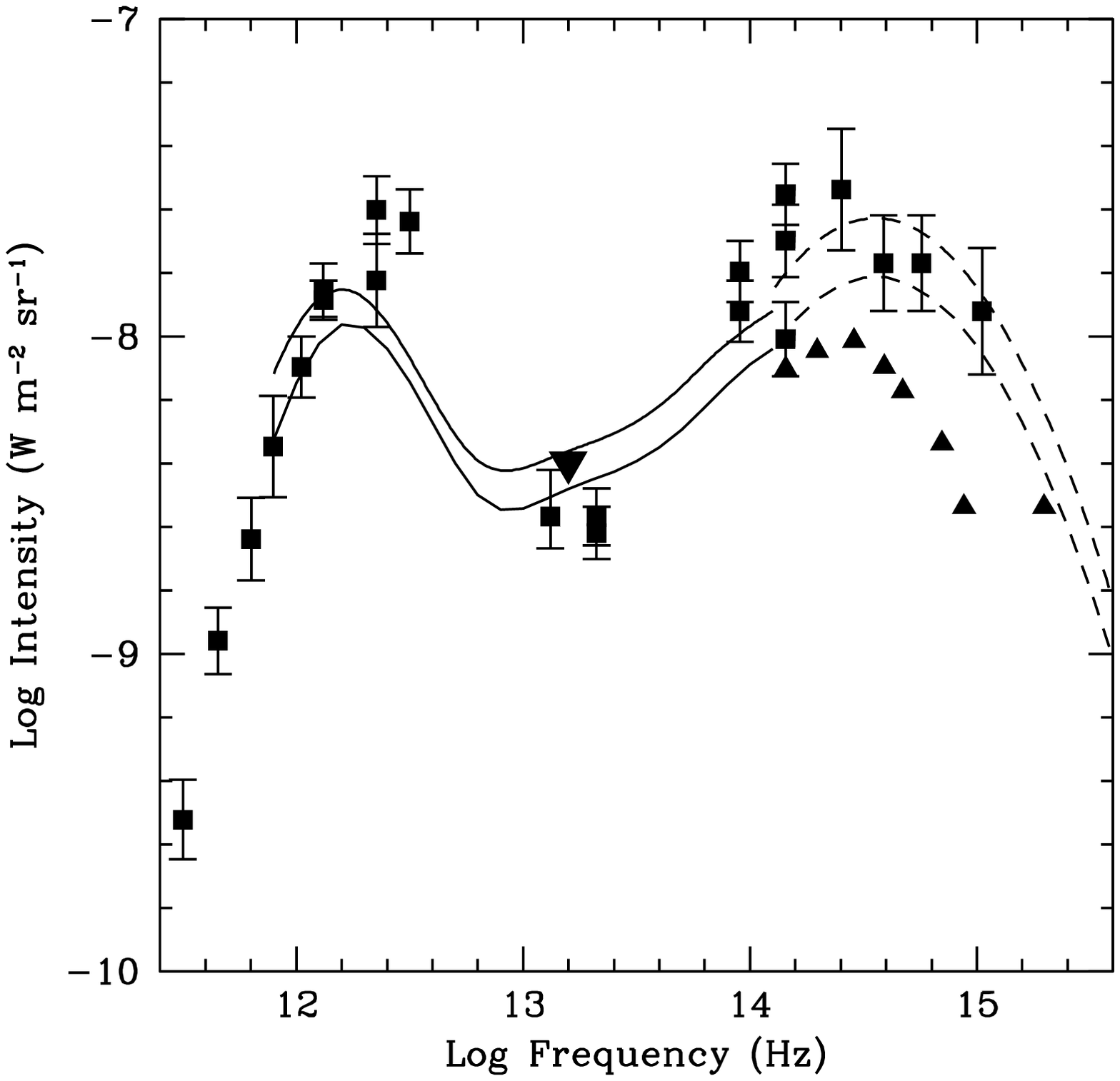}{3.4}{0.0}{-0.6}{
Spectral  energy distribution  of  the EBL,  taken  from SMS06.   Data
points  with  error bars  depict  measurements,  triangles show  lower
limits from source  number counts, and the inverted  triangle shows an
upper limit  from Stecker  and De Jager  (1997).  The upper  and lower
solid  curves  depict  FE  (fast  evolution) and  B  (baseline)  model
predictions. The dotted  lines show the extensions of  the models into
the optical--UV (from SMS06).
\label{fig:EBL}
}


\begin{deluxetable}{cccccc}

\tablecaption{
Blazar Spectral Indices in the 0.2 -- 2 TeV Energy Range and Isotropic
Luminosities at 1 TeV}
\tablewidth{0pt} 
\tablehead{ \colhead{Source} &
\colhead{$z$} & \colhead{$\Gamma_{obs}$} &
\colhead{$\Gamma_{s}$ (FE $\rightarrow$ B)} & 
\colhead{$\Gamma_{e}$ (FE $\rightarrow$ B)} & 
\colhead{$\cal{L}$(1 TeV) [$10^{36}$ W] } }
\startdata

1ES 2344+514 & 0.044 & 3.0 & 2.5 $\rightarrow$ 2.6 & 4.0 $\rightarrow$
4.2 &  2.9 \\  Mrk 180 &  0.045 &  3.3 & 2.9  $\rightarrow$ 3.0  & 4.8
$\rightarrow$  5.0  &  1.2  \\   1ES1959+650  &  0.047  &  2.7  &  2.3
$\rightarrow$  2.4 & 3.6  $\rightarrow$ 3.8  & 5.4  \\ PKS  2005-489 &
0.071 & 4.0  & 3.4 $\rightarrow$ 3.5 & 5.8 $\rightarrow$  6.0 & 8.6 \\
PKS 2155-304 & 0.117 & 3.3 & 2.2 $\rightarrow$ 2.4 & 3.4 $\rightarrow$
3.8 & 420  \\ H 2356-309 & 0.165  & 3.1 & 1.5 $\rightarrow$  1.9 & 2.0
$\rightarrow$  2.8  &  200  \\  1ES  1218+30  &  0.182  &  3.0  &  1.2
$\rightarrow$  1.6 & 1.4  $\rightarrow$ 2.2  & 310  \\ 1ES  1101-232 &
0.186 & 2.9  & 1.0 $\rightarrow$ 1.5 & 1.0 $\rightarrow$  2.0 & 230 \\
1ES 0347-121 & 0.188 & 3.1 & 1.2 $\rightarrow$ 1.7 & 1.4 $\rightarrow$
2.4 & 1200 \\ 1ES 1101+496 & 0.212 & 4.0 & 1.8 $\rightarrow$ 2.4 & 2.6
$\rightarrow$ 3.8 & 930 \\

\enddata
\vspace{-14pt}
\end{deluxetable}

\section{INTERGALACTIC ABSORPTION}
 \label{sec:absorption}

The   intergalactic    $\gamma$-ray   absorption   coefficient   (i.e.
optical
depth), $\tau(E,z)$, increases  monotonically with energy and
therefore
leads to  a steepening of  the intrinsic source  spectra as
observed at Earth. SMS06 give a useful parametric form for $\tau(E,z)$
with the
corrected parameters given in the erratum (Stecker, Malkan
\& Scully
2007). For sources at redshifts between 0.05 and 0.4,
Stecker  \&  Scully
(2006,  hereafter  SS06)  have  shown  that  this
steepening results  in a
well-defined increase in  the spectral index
of a source with an approximate power-law spectrum in the 0.2 -- 2 TeV
energy range with spectral  index $\Gamma_{obs}$.  This
increase is a
linear function in redshift $z$ of the form $\Delta \Gamma
= C + Dz$,
where   the    parameters   $C$   and   $D$    are   constants.    The
overall
normalization of  the source spectrum  is also reduced  by an
amount equal
to  $ exp \{-(A  + Bz)\}$, again  where $A$ and  $B$ are
constants. The
values of  $A, B, C,$ and $D$ are given  for the B and
FE  models in  SS06.
SS06 have  used this  relation to  calculate the
intrinsic  0.2 --  2  TeV
power-law $\gamma$-ray  spectra of  sources
having known redshifts in the
0.05 -- 0.4 redshift range for both the
B and FE models of EBL
evolution. A version of Table~2 of SS06 giving
values  for  intrinsic  spectral  index  of  the  source  $\Gamma_{s}$
is
shown  in Table  1.  Table  1  also shows  the respective  indices
$\Gamma_{e}  = 2\Gamma_{s}-1$
of  the electron  distributions  in the
sources under  the assumption that
the $\gamma$-rays  are produced by
inverse  Compton  interactions  in the
Thomson  regime.

Using  the
formula derived  in SS06,  we can estimate  the intrinsic
``isotropic
luminosity.''\footnote{We define isotropic  here as if the
source had
an  apparent  isotropic  luminosity  even though  blazars  are
highly
beamed  and   their  flux   (and  hence  their   apparent  luminosity)
is
dramatically enhanced  by relativistic Doppler  boosting.  This is
similar
to  the  nomenclature  used  for  $\gamma$-ray  bursts.   The
quantity $\cal{L}$ is equal to 4$\pi  \nu F_{\nu}$ given at h$\nu$ = 1
TeV in units of $10^{36}$  W ($10^{43}$ erg s$^{-1}$).}  The isotropic
luminosity
of the blazar  sources listed in Table 1  is obtained from
the formula
\begin{equation}
   {\cal{L}}     \simeq    4\pi    {{\Gamma_{o}-2}\over{\Gamma_{s}-2}}
(1+z)^{\Gamma_{s}-2} F_{o}[d(z)]^2 e^{(A+Bz)}
\end{equation}
where $d$ is the luminosity distance to the source, and $F_{o}$ is its
observed differential energy  flux at 1 TeV. The  other factors in the
equation give the k-correction  for the deabsorbed source spectrum and
the normalization correction factor for absorption given in SS06.  The
observational  references
for the sources  listed in  Table 1  can be
found in SS06 except for the new observations of 1ES 2344+514 (Albert
\etal 2007a), 1ES 1959 + 650 (Albert \etal 2006), 1ES 0347-121 (Aharonian
\etal 2007) and 1ES 1101+496 (Albert \etal 2007b). The source PG 1553+113 
at $z=0.36$ (included in Table~2 in SS06)  is not
listed here because the  
observations are in
the energy  range 0.09-0.6 TeV  and are therefore below  the operative
energy range  for applying the  analytic approximation given  in SS06.
The  blazars  Mrk 421  and  Mrk 501  are  not  included because  their
redshifts  are
significantly less than  0.05. However,  these blazars
are analysed by
Konopelko \etal (2003).

The numbers given in the last column of Table 1 are derived
for   the  fast
evolution   (FE)   model. One may note
that there  
appears to be a trend toward
blazars  having flatter  intrinsic
TeV spectra  and  higher isotropic
luminosities  at higher  redshifts.
However, one  must be  careful of
selection effects. The TeV photon fluxes
of these sources as observed
by {\it HESS} and {\it MAGIC}  only cover
a dynamic range of a factor
of  $\sim$20. Therefore,  only  brighter
sources can  be observed  at
higher  redshifts. This is  because of  both
diminution of  flux with
distance, and  intergalactic absorption (Stecker, de
Jager \& Salamon 1992). The observed
\teq{{\cal L}(z)} trend is naturally expected in a limited population sample
spanning a range of redshifts if the TeV-band fluxes are pegged
near an instrumental sensitivity threshold. A more powerful handle
on the intrinsic spectra and luminosities of these sources will be
afforded by the upcoming {\it GLAST} $\gamma$-ray mission, with its 
capability for detecting many blazars at energies below 200 GeV.

\section{PARTICLE DISTRIBUTIONS FROM SHOCK ACCELERATION}
 \label{sec:part_dist}

As  discussed  above,  inferences  of  source  spectra  from  specific
blazars
are  contingent  upon  the   particular  choice  of  an  EBL
model. Hence  there
is significant  uncertainty in deductions  of the
underlying  distribution
of emitting  electrons  in the  case of  SSC
models, or  protons in the case
of  hadronic models. It is  a goal of
this  presentation  to  provide  a
cohesive  connection  between  the
particle distributions in  blazars, the
resulting emission spectra in
the TeV  band, and the  spectrum of the EBL.

The  rapid variability
seen  in TeV  flares  drives the  prevailing  picture
for the  blazar
source  environment,  one  of  a compact,  relativistic  jet
that  is
structured on  small spatial scales that  are unresolvable by
present
$\gamma$-ray   telescopes.  Turbulence   in  the   supersonic  outflow
in
these jets naturally generates relativistic shocks, and these form
the
principal  sites  for  acceleration  of  electrons  and  ions  to
the
ultrarelativistic     energies     implied     by    the     TeV
$\gamma$-ray
observations. Within  the context of  this relativistic,
diffusive  shock
acceleration mechanism, numerical simulations are 
used here to derive expectations  for  the  energy distributions  of
particles  accelerated in
blazar  jets.

Diffusive  acceleration at
relativistic    shocks   is    less   well studied    than
that   for
nonrelativistic  flows, yet  it  is the  most applicable  process
for
extreme objects such as pulsar winds, jets in active galactic
nuclei,
and $\gamma$-ray bursts. Early work on relativistic shocks was
mostly
analytic  in the  test-particle approximation  ({\it e.g}., Peacock
1981,
Kirk \& Schneider 1987, Heavens \& Drury 1988), where the
accelerated
particles do not  contribute significantly to the global
hydrodynamic
structure  of  the  shock.  A key  characteristic  that
distinguishes
relativistic shocks from their non-relativistic
counterparts is their
inherent anisotropy due to  rapid convection of
particles through and
away downstream  of the shock. This  renders
analytic approaches more
difficult for  ultrarelativistic upstream flows,
though  advances can
be made in special cases,  such as the limit of
extremely small angle
scattering (pitch angle diffusion) (e.g. Kirk \&
Schneider 1987; Kirk
et al. 2000).  Accordingly, complementary Monte
Carlo techniques have
been  employed  for  relativistic  shocks  by  a  number
of  authors,
including test-particle analyses  for steady-state shocks of
parallel
and  oblique magnetic  fields by  Ellison, Jones  \& Reynolds
(1990),
Ostrowski  (1991), Bednarz  \& Ostrowski  (1998),  Baring (1999),
and
Ellison  \& Double  (2004). It  is such  a simulational  approach that
is
employed  here  to  illustrate  key spectral  characteristics  for
particles
accelerated  to high energies  at relativistic  shocks that
are  germane to
the  blazar emission-EBL  attenuation problem.  For a
recent  discussion  of
relativistic  shock acceleration,  see  Baring
(2004).

The   simulation   used   here   to   calculate   diffusive
acceleration in
relativistic  shocks is a Monte  Carlo technique that
has been  employed
extensively in supernova  remnant and heliospheric
contexts, and is
described  in detail in papers by  Ellison, Jones \&
Reynolds (1990),  Jones \&  Ellison (1991), Baring,
Ellison  \& Jones
(1994) and  Ellison,  Baring,  \&  Jones (1996).  It  is
conceptually
similar to  Bell's (1978)  test particle approach  to
diffusive shock
acceleration.  Particles   injected  upstream  gyrate   in  a
laminar
electromagnetic  field, and  particle trajectories  are determined
by
solving  a  relativistic  Lorentz  force  equation  in  the  frame  of
the
shock.  Because the  shock  is  moving with  a  velocity {\bf  u}
relative to the plasma rest frame, there will, in general, be a {\bf u
$\times$  B}   electric  field  in
addition  to   the  bulk  magnetic
field.  Alfv\'{e}n wave  turbulence is
modeled  by using a 
phenomenological description of ion
scattering  in  the  rest frame  of  the
plasma. The  scattering
allows particles to diffuse spatially along
magnetic field lines, and
to  varying extent,  across them  as well.  The
scatterings  are also
assumed to  be elastic,  an assumption that  is valid
so long  as the
flow  speed far  exceeds the  Alfv\'{e}n  speed. Hence,
contributions
from stochastic second-order Fermi acceleration, where the
scattering
centers move with the  Alfv\'{e}n waves, are generally
neglected. The
diffusion permits  a minority of  particles to cross  the
shock plane
numerous times, gaining energy with each crossing via
the shock drift
and first-order Fermi  processes.

A continuum of scattering angles,
between  large-angle  or small-angle
cases,  can  be  modeled by  the
simulation.  Denoting  local fluid  frame
quantities  by a  subscript
\teq{f},   the  time,   \teq{\delta   t_f},  between
scatterings   is
determined by  the mean free path, \teq{\lambda_f},  the
speed of the
particle,      \teq{v_f},     and      the      maximum     scattering
angle,
\teq{\thetascatt}, as  derived in Ellison,  Jones, \& Reynolds
(1990); for
small  angles,   it  is   given  by  \teq{   \delta  t_f=
\lambda_{f}
\thetascatt^{2}/(6v_f)},  a formula that  generally holds
to   within    10\%
for   \teq{\thetascatt   <    80^{\circ}}.   Here
\teq{\lambda_{f}} is
proportional to a power of the particle momentum
\teq{p} (e.g., see
Ellison  et al., 1990; Giacalone et  al., 1992 for
microphysical
justifications for this  choice), and for simplicity we
assume  that   it
scales  as  the   particle  gyroradius,  \teq{r_g},
i.e.    \teq{\lambda_{f}=\eta
r_{g}\propto    p}.

The    parameter
\teq{\eta}  in the  model is  a  measure of  the level  of
turbulence
present in the system, coupling directly to the amount of
cross-field
diffusion such  that \teq{\eta =1}  corresponds to the  isotropic Bohm
diffusion     limit.    It    can     be    related     to    parallel
(\teq{\kappa_{\parallel}=\lambda_{f}v/3})       and      perpendicular
(\teq{\kappa_{\perp}})  spatial  diffusion  coefficients  through  the
relation  \teq{\kappa_{\perp}/\kappa_{\parallel}=1/(1+\eta^{2})}  (see
Forman,  Jokipii \&  Owens, 1974,  Ellison et  al., 1995,  or Jokipii,
1987).   In the  parallel shocks  considered here,  where the  {\bf B}
field is
directed along the shock normal, \teq{\eta} has only limited
impact on
the resulting  energy spectrum, principally determining the
diffusive
scale normal to the shock. However, in oblique relativistic
shocks,  for
which the  field is  inclined to  the shock  normal, the
diffusive transport
of  particles across the field  (and hence across
the shock) becomes
critical to  retention of them in the acceleration
process.  Accordingly,
for such  systems, the  interplay  between the
field angle  and the value of
\teq{\eta} controls  the spectral index
of  the particle  distribution  (e.g. see
Ellison  and Double,  2004;
Baring, 2004).

\figureout{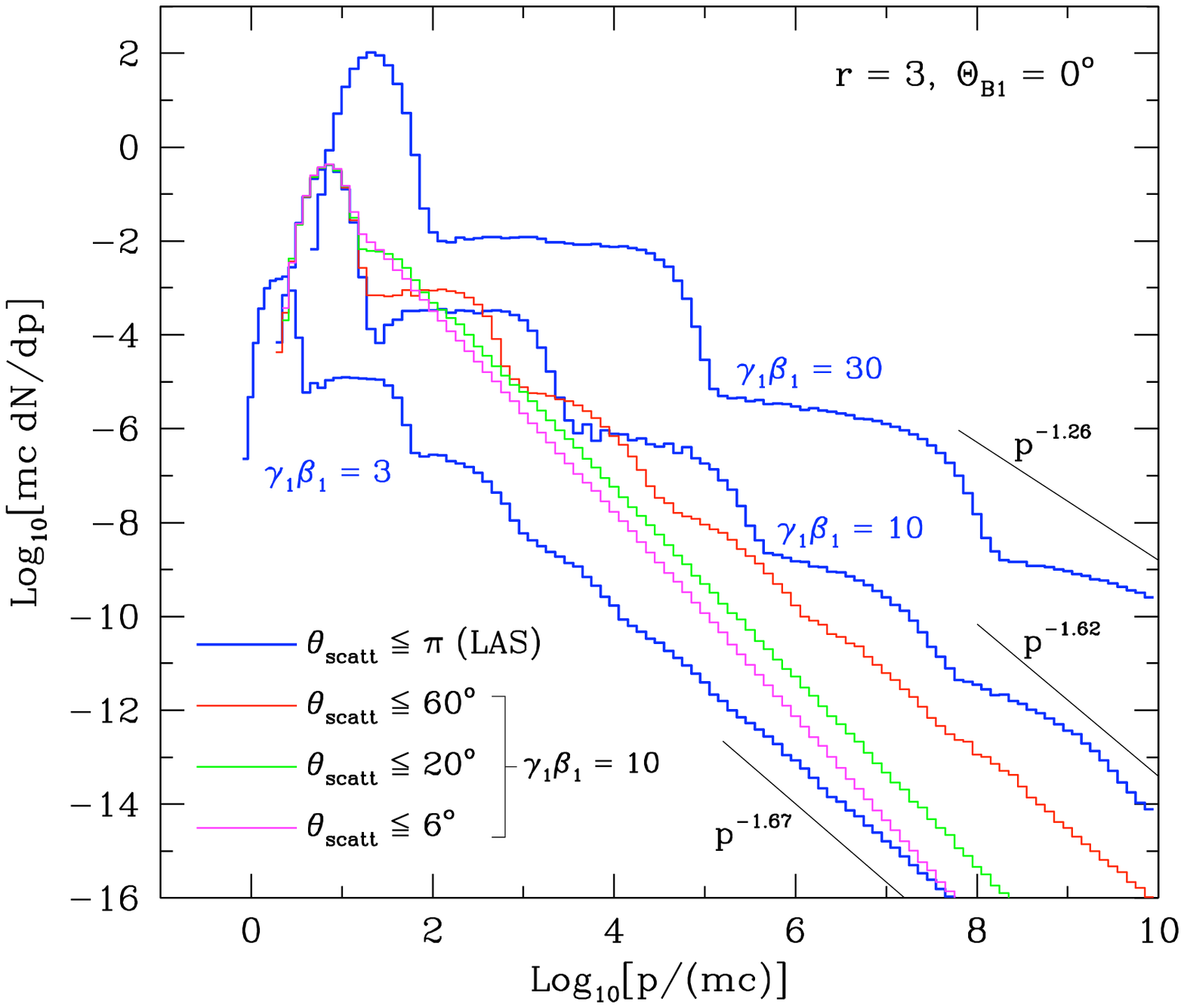}{3.7}{0.0}{-0.1}{
Particle       distribution      functions       \teq{dN/dp}      from
parallel
(\teq{\thetaBone=0^{\circ}}),    relativistic    shocks   of
upstream-to-downstream
velocity  compression ratio \teq{r=u_1/u_2=3},
as obtained  from a Monte
Carlo simulation of  particle diffusion and
gyrational  transport.  Three
shock  speeds  \teq{u_1=\beta_1 c}  are
depicted, namely
\teq{\beta_1=0.9487}, \teq{\beta_1=0.995} and
\teq{\beta_1=0.99944},
corresponding to the labels
\teq{\gamma_1\beta_1=3},
\teq{\gamma_1\beta_1=10} and
\teq{\gamma_1\beta_1=30}, respectively, on
the heavyweight
(blue)   histograms.   Scattering  off   hydromagnetic  turbulence
is
modeled by randomly deflecting  particle momenta by an angle
within a
cone, of  half-angle \teq{\thetascatt}, whose axis  coincides with the
particle
momentum prior  to scattering. The  heavyweight (blue) lines
are for the large angle scattering cases (LAS:
\teq{\thetascatt\leq\pi\gg
1/\gamma_1}), and these asymptotically
approach the  power-laws
\teq{dN/dp\propto p^{-\Gamma_{e}}} indicated
by lightweight lines, at
high and very high energies (not shown). For
the
\teq{\gamma_1\beta_1=10} case,  also exhibited are  three smaller
angle
scattering     cases,    \teq{\thetascatt     \leq    60^\circ}
(red),
\teq{\thetascatt \leq 20^\circ} (green), and \teq{\thetascatt
\leq
6^\circ} (magenta) corresponding to pitch angle diffusion
(PAD).  These
have high-energy asymptotic power-law indices of
\teq{\Gamma_{e} =1.65},
\teq{\Gamma_{e} =1.99} and \teq{\Gamma_{e}
=2.20}, respectively.\\
\label{fig:laspad}
}

Representative particle distributions  that result from our simulation
of
diffusive  acceleration  at relativistic  shocks  are depicted  in
Figure~\ref{fig:laspad},
highlighting  several  key features.   These
distributions are equally
applicable to electrons or ions, and so the
mass  scale is  not specified.   The spectral  index declines  and the
distribution is  flatter for
faster shocks with  larger upstream flow
(bulk) Lorentz  factor
\teq{\gamma_1}, when the  velocity compression
ratio  \teq{r}  is fixed.
This  is  a  consequence  of the  increased
kinematic energy  boosting
occurring at relativistic  shocks.  Such a
characteristic  is  evident,  for
example  in  the work  of  Kirk  \&
Schneider (1987),  Ballard \& Heavens
(1991)  and Kirk et  al. (2000)
for  the  case  of  pitch  angle scattering,
and  Ellison,  Jones  \&
Reynolds (1990), Baring (1999)  and Ellison \&
Double (2004) for much
larger   angle   scattering.    What   is   much   more
striking   in
Figure~\ref{fig:laspad}   is  that   the  slope   and  shape   of  the
nonthermal
particle  distribution  depends   on  the  nature  of  the
scattering.  The
asymptotic, ultrarelativistic index of
\teq{\Gamma_{e} =2.23} is
realized only in the mathematical limit of
small (pitch)  angle diffusion
(PAD), where the  particle momentum is
stochastically deflected on
arbitrarily  small angular (and therefore
temporal)  scales.    In  practice,
PAD  results   when  the  maximum
scattering  angle \teq{\thetascatt} is
inferior  to the  Lorentz cone
angle  \teq{1/\gamma_1}  in  the  upstream
region.   In  such  cases,
particles  diffuse in  the region  upstream of  the
shock  only until
their angle to the shock normal exceeds around
\teq{1/\gamma_1}. Then
they  are rapidly  swept to  the downstream  side of
the  shock.  The
energy gain per shock crossing cycle is then roughly a
factor of two,
simply  derived from  relativistic kinematics  (Gallant \&
Achterberg
1999;   Baring   1999).

To    contrast   these   power-law   cases,
Figure~\ref{fig:laspad}  also  shows   our  results  for
large  angle
scattering scenarios  (LAS, with \teq{\thetascatt\sim\pi}),
where the
spectrum   is  highly   structured   and  much   flatter  on   average
than
\teq{p^{-2}}. The structure,  which becomes extremely pronounced
for
large  \teq{\gamma_1},  is   kinematic  in  origin,  where  large
angle
deflections lead to the distribution of fractional energy gains
between
unity  and \teq{\gamma_1^2} in  successive shock  transits by
particles.
Gains like this are  kinematically analogous to photon
energy
boosting by Compton scattering.  Each structured bump
or
spectral segment  shown in Figure~\ref{fig:laspad}  corresponds to
an increment in the
number of shock crossings, successively from
\teq{1\to3\to 5\to 7} etc.,
as illustrated by Baring (1999); they
eventually smooth  out to
asympotically approach  power-laws that are
indicated by the lightweight
lines in the Figure. {\it The indices of
these  asymptotic  results  are
all  in  the  range}  \teq{\Gamma_{e}
<2}. Intermediate cases are also
depicted in Figure~\ref{fig:laspad},
with \teq{\thetascatt\sim  4/\gamma_1}. The
spectrum is  smooth, like
the  PAD case,  but the  index is  lower  than 2.23.
Astrophysically,
there is no reason to  exclude such cases.  From the plasma
point  of view,  magnetic  turbulence could  easily be
sufficient  to
effect scatterings  on this  intermediate angular scale,  a contention
that becomes
even more salient for ultrarelativistic shocks with
\teq{\gamma_1\gg
10}.  It is also evident that a range of spectral
indices is produced
when \teq{\thetascatt} is of the order of
\teq{1/\gamma_1}, In this
case, the scattering processes corresponds
to a transition between  the
PAD and LAS limits.

Given the results
of our  numerical simulations, the  implications for
distributions of
relativistic particles in blazars  are apparent. There
can be a large
range  in  the  spectral  indices \teq{\Gamma_{e}}  of  the
particles
accelerated in relativistic  shocks, and these indices usually
differ
from \teq{\Gamma_e\sim 2.23}.  They can be much steeper,
particularly
in  oblique  shocks  (e.g.,  Ellison \&  Double  2004;  Baring
2004).
However, they can also  be much flatter, so that
quasi-power-law
particle distributions  \teq{p^{-\Gamma_{e}}} with
\teq{\Gamma_{e}} $
\le  2$   are  readily  achievable.    Such  flat
distributions  from
relativistic shock  acceleration have not  usually been
admitted when
considering  properties of  blazar  jets and  their possible  emission
spectra.\footnote{It has been suggested
that electron distributions  with $\Gamma_{e} < 2$ can  be obtained by
stochastic  acceleration in combination  with boundary  layer particle
trapping which produces a pileup effect (Ostrowsky 2000).}\footnote{Our
results require the  implicit assumption that the
electron spectra produced during blazar flares are not significantly 
affected by cooling by synchroton radiation. This requires that 
$t_{acc} < t_{ccol}$ which constrains both the magnetic field strength 
and the electron energy (Baring 2002).}

\section{CONCLUSIONS}

This   finding   that   relativistic   shock   acceleration   produces
particle spectra  with  a  significant  range  of  spectral  indices,
including  those with  $\Gamma_{e} \le  2$  corresponding to  inverse
Compton $\gamma$-ray spectra  with $\Gamma_{s} \le  1.5$, has various
consequences. The considerable diversity in the values
\teq{\Gamma_{e}} produced in
relativistic shocks is matched by the
diversity in  the intrinsic
spectral indices of  blazars indicated in
Table 1. Moreover, particle  distributions with $\Gamma_{e} \le 2$ are
consistent with the inferred values for the three most distant blazars
listed in  the table.  A hard TeV  $\gamma$-ray spectrum with  a value
$\Gamma_{s} < 2$  within the context of SSC  model building (see, {\it
e.g.}, Stecker, De  Jager \& Salamon 1996), indicates  that the energy
range  of the  observation is  below the  Compton peak  energy  in the
spectral energy  distribution of the source, which  is given by
$E^2$
times   the  differential  photon   spectrum.  For   extreme  blazars,
this
peak can easily be at an energy above 2 TeV (de Jager \& Stecker
2002).  A simple SSC model prediction then follows. The observation of
an approximate  power-law spectrum in the sub-TeV  energy range should
imply approximately the same index  as the synchrotron emission in the
optical to X-ray band.

Our  reexamination of  blazar  $\gamma$-ray spectra  in  the light  of
relativistic shock acceleration theory has important implications
for
constraining the flux  of the EBL in the  near infrared. Specifically,
the  low  values  of  $\Gamma_{e}  \le  2$  readily  obtained  in  our
numerical
results implies an increase in the upper limit on the
near
infrared EBL to values above  that obtained by Aharonian \etal (2006).
Such  a
result is  consistent with  the fast  galaxy  evolution model
which appears to be favored by the {\it Spitzer} observations.

Table 1  hints at
a  redshift evolution of  TeV blazars with  a trend
toward  flatter spectra
and higher  isotropic luminosities  at higher
redshifts. Although  only ten blazars are  listed in the  table and
selection  effects are
important,  one  may speculate  as to  whether
there is a general trend in
blazar activity in the form of higher jet
Doppler factors at higher
redshifts.  Future combined {\it GLAST}-TeV
broadband   spectral  data  will   further  define   intrinsic  source
properties,  thus  enabling  the  further  investigation  of  possible
redshift evolution  of blazar  flux and spectral  characteristics. The
{\it H.E.S.S.} and {\it MAGIC} atmospheric
\v{C}erenkov TeV telescopes have had remarkable success in observing
blazars.  With  the {\it VERITAS} atmospheric  Cherenkov telescope now
on-line,  the  population  of  known  TeV  blazars  will  be  extended
considerably.

\section*{acknowledgments}
We wish to thank the referee for helpful comments.
MGB and EJS acknowledge the support of NASA through Grant 
No.~NNG05GD42G and the National Science Foundation under 
Grant No.~AST00-98705 for parts of this research. 

\end{document}